# Realization of exciton-mediated optical spin-orbit interaction in organic microcrystalline resonators

Jiahuan Ren,[1,2] Qing Liao,[1,*] Xuekai Ma,[3,*] Stefan Schumacher,[3,4] Jiannian Yao,[2] Hongbing Fu[1,*]

[1]Beijing Key Laboratory for Optical Materials and Photonic Devices, Department of Chemistry, Capital Normal University, Beijing 100048, People's Republic of China

[2] Tianjin Key Laboratory of Molecular Optoelectronic Science, School of Chemical Engineering and Technology, Tianjin University and Collaborative Innovation Center of Chemical Science and Engineering (Tianjin), Tianjin 300072, P. R. China

[3]Department of Physics and Center for Optoelectronics and Photonics Paderborn (CeOPP), Universität Paderborn, Warburger Strasse 100, 33098 Paderborn, Germany

[4]Wyant College of Optical Sciences, University of Arizona, Tucson, Arizona 85721, United States

**Abstract:** The ability to control the spin-orbit interaction of light in optical microresonators is of fundamental importance for future photonics. Organic microcrystals, due to their giant optical anisotropy, play a crucial role in spin-optics and topological photonics. Here we realize controllable and wavelength-dependent Rashba-Dresselhaus spin-orbit interaction, attributed to the anisotropic excitonic response in an optical microcavity filled with an organic microcrystalline. We also investigate the transition of the spin-orbit interaction from dominant photonic type caused by the splitting of the transverse-electric and transverse-magnetic modes to spin-orbit interaction of the Rashba-Dresselhaus type. The interplay of the two allows us to engineer the spin-orbit interaction of light in organic microcavities, which besides its fundamental interest promises applications in spin-controlled on-chip integrated nanophotonic elements, towards exploiting non-magnetic and low-cost spin-photonic devices.

## INTRODUCTION

The spin degree of freedom of electrons has been extensively exploited in condensed matter physics and is at the heart of many protocols for efficient data transfer, quantum-information processing, and storage.(*1, 2*) The fundamental physics of their potential applications relies on the splitting of the electronic energy bands induced by spin-orbit interaction (SOI), which is a relativistic effect caused by the interaction between moving electrons and magnetic fields.(*3*) The SOI of electrons in solid-state systems was originally realized in structure-inversion-asymmetric materials and bulk-inversion-asymmetric materials by Rashba(*4*) and Dresselhaus,(*5*) respectively. Light possesses the intrinsic polarization degree of freedom, where the right- and left-handed circular polarizations are akin to the spin characters of electrons. In optics, the SOI of light has opened remarkable new opportunities to manipulate the pseudospin of photons,(*6, 7*) which has inspired the study of the spin Hall effect of light (SHEL)(*8-11*) and optical topological insulators.(*12-14*) In these areas, the synthetic gauge fields created by the SOI play a central role.

It is known that the SOI of light in optical microcavities can be caused by the transverse electric-transverse magnetic (TE-TM) mode splitting(*15, 16*), which can be modeled as an effective magnetic field and by constructing artificial gauge fields.(*17*) Besides the TE-TM splitting, the SOI in optical microcavities can also be induced by other effects, such as the chirality of the materials as well as the so-called Rashba-Dresselhaus (RD) Hamiltonian, which can also contribute to the construction of the synthetic gauge fields to further manipulate the spin states of photons. It is known that the effective magnetic field caused by the TE-TM splitting winds twice in reciprocal space,(*9*) while the effective magnetic field caused by the RD effect winds only once.(*18*) The physical reason is that the TE-TM splitting preserves the spatial inversion symmetry, in contrast to the Rashba or Dresselhaus SOI.(*5, 19, 20*) It remains a great challenge but is highly desirable to integrate multiple of these different types of SOI into the same optical microcavity. This would not only facilitate the tuning of the spin splitting between different types but also provide an interesting platform for the study of a novel type of topological photonics to efficiently control the pseudospin of light.

In the study of the SOI of light, organic crystalline materials have received significant attention due to their large optical anisotropy. Recently, the controllable spin splitting and observation of different types of SOI have been reported in optical microcavities filled with liquid crystals (LC)(*21*) and α-perylene single-microcrystals.(*22*) In the LC microcavities, the TE-TM splitting is insignificant and the RD SOI dominates. In the α-perylene single-microcrystal microcavities, the SOI arises because of the TE-TM splitting, but it is difficult to realize the RD effect due to the nearly constant energy splitting $\beta_0$ of the linearly polarized modes. This energy splitting $\beta_0$ in the α-perylene single-microcrystal microcavities results mainly from the non-resonant linear birefringence of the anisotropic organic crystals, with a much smaller contribution from the excitonic response, that is not explicitly utilized or discussed in detail. Tuning the optical modes more into resonance with the excitonic excitations, however, introduces a pronounced exciton-mediated contribution to the system's SOI with fundamental changes to the topology of the overall optical response: i) Due to the resonant nature the energy splitting $\beta_0$ becomes wavelength-dependent instead of a constant. In other words, the excitons only strongly influence those cavity photon modes that are near the resonance. As the cavity modes move away from the excitonic resonance, the influence of the excitons reduces gradually. ii) With the pronounced anisotropy of the dipole coupling to the excitonic excitations, coupling only occurs to one of the two perpendicular cavity modes, leaving the other one unaffected.(*23*) With these fundamental observations, involving the excitonic response brings additional controllable degrees of freedom enabling us to introduce and tailor different types of SOI in the same system, which may lead to interesting applications in topological photonics and quantum phase transitions.(*24*)

In the present work, by regulating the exciton-photon interaction, we artificially engineer the SOI and realize the coexistence of photonic TE-TM splitting and the RD effect in the same microcavity system. In our experiments, the organic microcavity is filled with a thick β-phase single-crystalline perylene. Due to the moderate crystalline anisotropy, in the vicinity of the exciton resonance, the curvatures of the TM modes become much smaller, while the TE modes are not substantially influenced. This leads to the possibility for one TE mode to simultaneously cross with multiple TM modes and allows the introduction of photonic and RD SOIs in the same microcavity. The transition of the spin splittings induced by the different types of SOIs is investigated in detail. With this we introduce an approach not relying on strong external magnetic fields that by using purely optical means allows the SOI control in organic microcrystalline resonators, which gives access to a unique toolbox for advanced quantum control strategies in on-chip integrated photonics.

## RESULTS AND DISCUSSION

### Principles of spin-orbit interaction in optical microcavities

To realize complex SOIs in one microcavity, we consider an anisotropic cavity medium which can possess TE-TM splitting, linear birefringence, as well as resonant exciton-photon interaction. Generally, the TE-TM splitting in planar microcavities increases with the in-plane momentum (proportional to $k^2$ with $k$ being the in-plane momentum) and vanishes at $k = 0$.(16) The contribution of the linear birefringence can lead to the energy splitting at $k = 0$. Due to the coupling to anisotropic excitons, the energy splitting that occurs at $k = 0$ becomes wavelength-dependent and only one of the two orthogonally polarized cavity modes is significantly influenced. As a result, the strongly affected cavity modes start to move away from their counterparts of the same parity and towards their counterparts of opposite parity. When two perpendicular cavity modes of opposite parity meet each other at $k = 0$, the RD SOI is induced, accompanied by a spin splitting. The spin splitting caused by the RD effect is proportional to $k$. In such microcavity, in a simplified picture the system's effective Hamiltonian ($2 \times 2$) in the circular polarization basis can be written in the form,(21, 25)

$$H(\mathbf{k}) = \begin{pmatrix} \frac{\hbar^2}{2m_x} + \frac{\hbar^2}{2m_y} + \alpha k_y & \beta_0 + \beta_1 \mathbf{k}^2 e^{2i\varphi} \\ \beta_0 + \beta_1 \mathbf{k}^2 e^{-2i\varphi} & \frac{\hbar^2}{2m_x} + \frac{\hbar^2}{2m_y} - \alpha k_y \end{pmatrix} \quad (1)$$

where $m_x$ ($m_y$) is the effective mass of cavity photons along $x$ ($y$) direction, which are typically different in anisotropic media. $\alpha$ represents the strength of the linear splitting or RD effect (assuming the Rashba and Dresselhaus coupling strengths equal to each other)(21), contributing to only $k_y$ direction in the in-plane momentum ($k_x, k_y$) space. $\beta_1$ represents the strength of the TE-TM splitting, and $\varphi$ ($\varphi \in [0, 2\pi]$) is the polar angle. $\beta_0$ denotes the energy splitting of linear polarized modes of opposite parity at $k = 0$ (here, we define it as $\beta_0 = E_{\text{TM}} - E_{\text{TE}}$, where $E_{\text{TM}}$ and $E_{\text{TE}}$ are the ground state energies of TE and TM modes of opposite parity). Different from the α-perylene crystal,(22) the value of $\beta_0$ in this case is determined by two factors, i.e., the linear birefringence and the anisotropic exciton-photon interaction. The contribution from the linear birefringence is linked to the crystallographic axes and thus remains constant due to the immutability of organic crystals.(23) However, due to the wavelength-dependent anisotropic exciton-photon interaction, the value of $\beta_0$ varies gradually with the wavelength of the cavity modes.

We first consider the case of $\beta_0 < 0$. Two eigenstates of the Hamiltonian (1) are shown in Fig. 1A-D with $\alpha \neq 0$, $\beta_1 \neq 0$ and $m_y = 1.5 \, m_x$. Here, the two perpendicularly polarized cavity modes have opposite parity. The strong TE-TM splitting and anisotropy result in that the eigenmodes in reciprocal space are strongly squeezed (not circles anymore) along $k_x$ and $k_y$ directions (Fig. 1C, where two perpendicular ovals are formed), respectively. The lower eigenmode in Fig. 1B is flatter compared with the upper one, while the situation reverses in Fig. D where the two modes cross at finite momentum. These kinds of crossing points cannot be observed in Fig. D, because the linear splitting, which is the determining factor to open them, lies only along $k_y$ direction, without contribution to $k_x$ direction. When $\beta_0 = 0$, the two cavity modes are in resonance at $k = 0$, leading to the spin splitting (i.e., the RD effect) along $k_y$ direction as shown in Fig. 1E. When $\beta_0 > 0$, the linear splitting results in the anti-crossing of the two cavity modes (Fig. 1F), which was observed in our previous report.(*22*) According to the above analysis, the strong anisotropy of the material and the mediation of the excitons may give rise to the intersecting of a TE (TM) mode with multiple TM (TE) modes (c.f. Fig. 1G). This configuration requires more cavity photon modes with small energy difference between them, which can be achieved in a relatively thicker microcavity. In this case, the anti-crossing at finite momentum and the RD effect close to zero momentum may occur simultaneously as predicted in Fig. 1H (note that the dispersion relation presented in Fig. 1H is a schematic diagram, not calculated directly from a $4 \times 4$ Hamiltonian).

**Experimental realization**

To experimentally realize the above theoretical prediction, we fabricate an optical microcavity, in which β-phase single-crystalline perylene nanosheets are sandwiched between two-layer metallic films as sketched in Fig. 2A. Perylenes, due to their excellent optical properties,(*26, 27*) special polaritonic bands, and birefringent crystals,(*18*) provide a wonderful platform for the study of topological photonics.(*22*) More importantly, perylene polymorphs provide an effective approach that affects the physicochemical properties of crystals and avoid the difficulty in time-consuming molecular synthesis.(*28, 29*) The microscope image clearly shows that these nanosheets have uniform rhombus morphology and smooth surface (Fig. 2B). The X-ray diffraction (XRD) analysis indicates that these rhombus sheets possess the β-phase structure (Fig. S1), i.e., their unit cell comprises two molecules and forms a standard herringbone staking pattern (Fig. 2C). The characteristic photon stopband in the transmission spectrum has also confirmed this assignment (Fig. S2).

We perform angle-resolved reflectivity (ARR) measurements of the microcavity with 3.2-μm β-phase perylene (Fig. S3) at room temperature. The thicker microcavity gives rise to the relatively small energy difference between neighboring cavity modes, which is essential to observe the crossing of multiple cavity modes. Fig. 2D shows the corresponding two-dimensional (2D) reflectivity dispersion at $k_x = 0$ (i.e., along Y-direction as indicated in Fig. 2C) and two sets of modes with different curvatures can be clearly seen. One set of the modes marked by the red dotted lines has larger curvatures, which is almost wavelength-independent and evenly distributed in the spectral range between 400 nm and 660 nm. These modes are consistent with the fitted individual TE cavity modes which are calculated by using the coupled harmonic oscillator (CHO) model.(*30*) The other set of the modes (TM modes) marked by the yellow dotted lines exhibits wavelength-dependent curvatures, i.e., they become smaller as approaching the exciton resonance of 469 nm. Note that the experimentally measured TM modes agree well with the calculated ones (considering the participation of excitons) at larger momenta, but they bifurcate at smaller momenta because of the RD effect (see the detailed discussion below). We further measured the polarization-dependent ARR of the cavity modes by adding a linear polarizer to the detection optical path (Fig. S4). It is found that these two sets of the cavity modes are perfectly orthogonally polarized, i.e., TE modes are polarized along X-direction, whereas TM modes along Y-direction. Obviously, the wavelength-dependent curvatures of TM modes are a typical feature given by the exciton-photon interaction. The influence of the excitons on TE modes, however, can be neglected. The anisotropic light-matter interaction enables the intersection of a TE mode with multiple exciton-mediated TM modes, towards the realization of the theoretical hypothesis in Fig. 1H.

**Interaction of multiple cavity modes**

Fig. 3A shows the magnified ARR spectrum (Fig. 2D). Clearly, the energy splitting $\beta_0'$ (defined as $\beta_0' = E_{TM}^n - E_{TE}^{n-1}$, where the superscript n represents the mode index) reduces gradually and finally approaches 0 around 575 nm (see also Fig. 4A). The non-constant splitting $\beta_0'$ results in complicated and rich mode interaction in such a microcavity. Firstly, the adjacent TE and TM modes of the same parity cross at $|k| > 0$, for example, clear crossing points appear at 496 nm (black dashed circle). Secondly, TE and TM modes of opposite parity anticross at around $k_y = \pm 10.15$ μm$^{-1}$ (white dashed circles), similar to the phenomenon observed in the α-phase-perylene microcavity.(*22*) This anti-crossing section is validated by the 2D tomography measurements (Fig. 3B, more details can be found in Fig. S5). It is worth mentioning that the anticrossings occur for the TM$_n$ mode and the TE$_{n+1}$ mode, hence it corresponds to the case of $\beta_0 > 0$ as shown in Fig. 1F. Importantly, at longer wavelengths, TE and TM modes of opposite parity are close to each other and eventually present energy degeneracy at $k = 0$ (blue dashed circle). Comparing the experimental and simulated (yellow lines in Fig. 2D) results of the TM modes, one can see their obvious deviation at $k_y = 0$. At the same time, different spin components split along $k_y$ directions (see Fig. 3K and Fig. S6), evidencing the RD effect. If focusing on the cavity TE$_4$ mode in Fig. 3A, it is clear that it anticrosses with the TM$_3$ mode at larger momentum and almost interacts with the TM$_5$ mode at $k = 0$, which triggers the RD spin splitting (Fig. 3J). This behavior is consistent with our theoretical expectation in Fig. 1H.

**Spin splitting and transition**

We performed three-dimensional (3D) tomography and measured the Stokes vector components(*31*) in energy-momentum space to analyze the spin splitting. From the 2D wavevector map of the tomography in Fig. 3B, we can clearly see two mutually perpendicular and independent ovals that represent the TE$_2$ (the inner most horizontal oval marked by the dashed green line) and TM$_3$ (the vertical oval marked by the dashed red line) cavity modes, respectively. Additionally, the TM$_3$ mode crosses with the TE$_3$ mode (the mode outside of the TE$_2$ mode) and anticrosses with the TE$_4$ mode (the mode outside of the TE$_3$ mode). Around the anti-crossing point at larger momentum ($k_y \approx 10.15$ μm$^{-1}$), the photonic SOI caused by the TE-TM splitting dominates. Consequently, the SHEL can be observed (dashed circle region in S$_3$ component of the Stokes vector in Fig. 3G). We note that at this wavelength the RD SOI cannot be clearly recognized. As the wavelength increases, TM modes shrink faster than TE modes in reciprocal space. At some point (e.g., white dashed line in Fig. 3A), one can see that the TM$_4$ mode becomes tangent to the outer TE$_4$ mode along $k_y$ direction and simultaneously is tangent to the inner TE$_3$ mode along $k_x$ direction (Fig. 3C). Here, the near resonant TE$_3$ and TM$_4$ modes enhances the RD effect, leading to the partial spin splitting (i.e., the spin components of the two innermost circles are not completely split) as shown in Fig. 3H. At the wavelength 532 nm (blue dashed line in Fig. 3A), although the TE$_3$ mode disappears, the TM$_4$ mode is still visible (Fig. 3D) and the spin components strongly split as can be seen in Fig. 3I. Remarkably, in Fig. 3I, SHEL with twice winding of the TE$_4$ mode (the mode outside of the innermost one) is still visible although the contrast becomes weaker in comparison with the TE$_4$ mode (white dashed circle) in Fig. 3G. However, as the wavelength increases further to 542 nm, corresponding to the cyan dashed line in Fig. 3A, the spin splitting and single winding of the TE$_4$ mode can be clearly seen (innermost ring in Fig. 3J). The detailed transition process between the states in Fig. 3I and 3J can be found in Fig. S7. From Fig. 3E, one can see that there is no exchange between the $TE_4$ mode and the outer $TM_5$ mode, indicating that the RD effect is not dominating here. The spin splitting caused by the RD SOI can be distinctly observed at round 550 nm (Fig. 3F and 3K). For an even longer wavelength (~600 nm), two orthogonally polarized modes of opposite parity ($TE_6$ and $TM_7$) are brought into resonance with $\beta'_0 = 0$ (Fig. 4A) and a very clear splitting of the parabolic dispersions along $k_y$ direction can be seen in Fig. 3A (spin-dependent dispersions can be found in Fig. S6).

Figures 4B-4D present six polarization components, collected at $k_y = 3$ μm$^{-1}$ along the vertical black dashed line in Fig. 3A, depending on the mode indices (or the wavelength). The horizontally and vertically polarized components ($S_1$ in Fig. 4B) become weaker at larger indices (longer wavelength), while the trend of the circularly polarized components ($S_3$ in Fig. 4D) is opposite, demonstrating that the RD SOI dominates at longer wavelength where $\beta'_0 \to 0$. Interestingly, the diagonally and antidiagonally polarized $S_2$ components swap sign around the transition region of different SOIs (Fig. 4C).

**Conclusion**

In conclusion, we have realized both photonic and RD SOIs in a β-phase single-crystalline perylene (organic) microcavity, mediated by anisotropic and frequency dependent exciton-photon interaction. In such microcavity, the exciton dipoles oriented along Y-direction affect only the cavity TM modes, which, together with the TE-TM splitting, leads to the dispersion of the TM modes to have much smaller curvature. Consequently, we observed simultaneously the spin splitting carrying a double winding, induced by the TE-TM splitting, at larger momenta and the spin splitting carrying a single winding, induced by the RD SOI, at smaller momenta. Different spin splittings can even appear on the same cavity mode. Hence this system provides a platform for the study of the transition of the SOI from the photonic type to the RD type. Our work also paves the way for studying the influence of the RD SOI on the strong light-matter coupling states, e.g., exciton-polaritons, in semiconductor microcavities and manufacturing low-cost spin photonic devices.

**Acknowledgments:**

**Funding:** This work was supported by the National Key R&D Program of China (Grant No. 2018YFA0704805, 2018YFA0704802 and 2017YFA0204503), the National Natural Science Foundation of China (22090022, 21833005, 21873065, 21790364 and 21673144), the High-level Teachers in Beijing Municipal Universities in the Period of 13th Five-year Plan (IDHT20180517 and CIT&TCD20180331), Beijing Talents Project (2019A23), the Open Fund of the State Key Laboratory of Integrated Optoelectronics (IOSKL2019KF01), Capacity Building for Sci-Tech Innovation-Fundamental Scientific Research Funds, Beijing Advanced Innovation Center for Imaging Theory and Technology. The Paderborn group acknowledges the support from the Deutsche Forschungsgemeinschaft (DFG) through the collaborative research center TRR142 (grant No. 231447078, project A04) and Heisenberg program (grant No. 270619725).

**Author contributions:** J.-H.R. and Q.L. designed the experiments and performed experimental measurements. X.M. and S.S. performed the theoretical calculation and analysis. J.-H.R., Q.L. and X.M. wrote the manuscript with contributions from all authors. Q.L., J.-N.Y. and H.-B.F. supervised the project. All authors analyzed the data and discussed the results.

**Competing interests:** Authors declare no competing interests.

**Data and materials availability:** All data needed to evaluate the conclusions in the paper are presented in the paper and/or the Supplementary Materials. Additional data related to this paper may be requested from the authors.

**Corresponding authors:** Q.L.: liaoqing@cnu.edu.cn; X.M.: xuekai.ma@gmail.com; H.-B.F.: hongbing.fu@tju.edu.cn

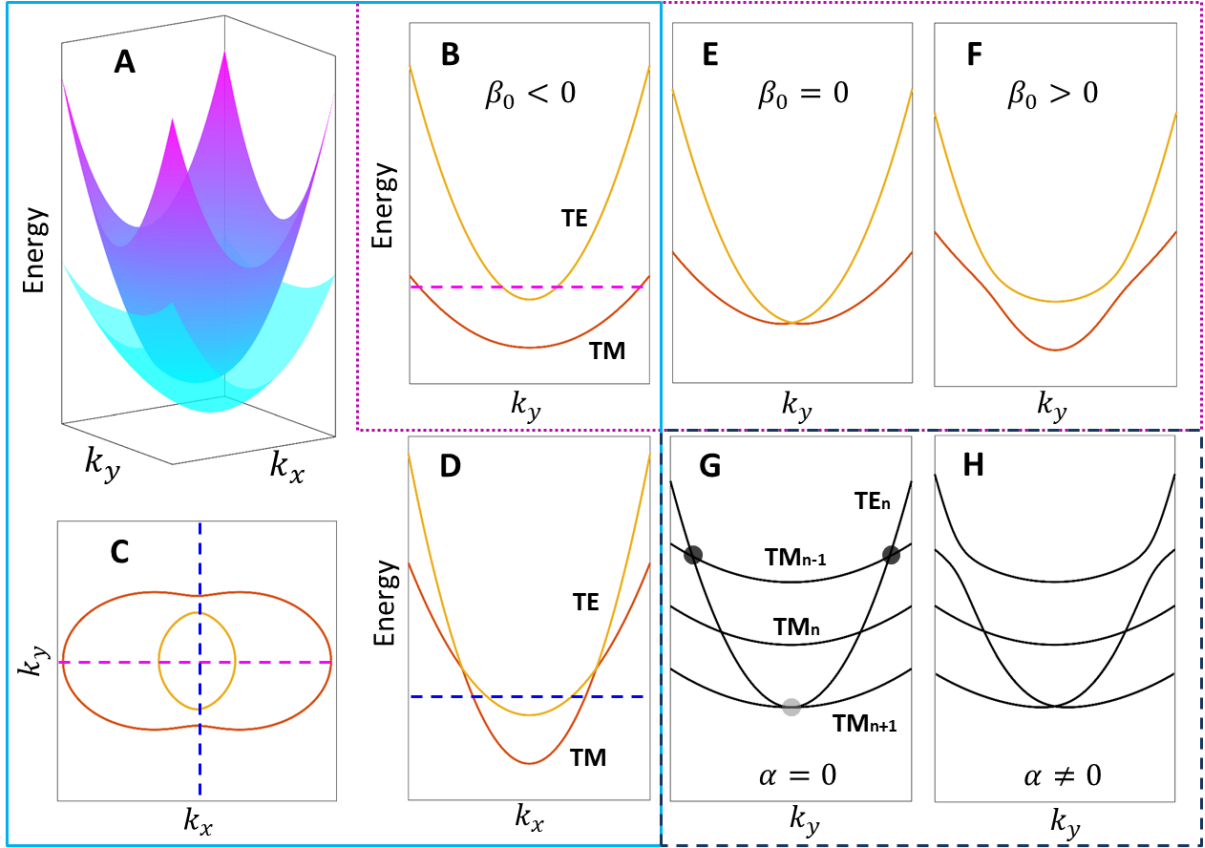

**Fig. 1. Spin orbit interaction of light in an optical microcavity.** (A) 3D dispersion calculated form the Hamiltonian (1) with $m_y = 1.5\ m_x$ and $\beta_0 < 0$. The two cavity modes are perpendicularly polarized (TE and TM) and have opposite parity. (B,D) Cross sections of the dispersion in (A) at $k_x = 0$ (B) and $k_y = 0$ (D). (C) Cross section (in $k_x - k_y$ plane) of (A) at the energy corresponding to the dashed lines in (B,D). (E,F) Dispersions calculated from the Hamiltonian (1) with $\beta_0 = 0$ (E) and $\beta_0 > 0$ (F), and the other parameters are the same with those in (B). (G,H)

Schematic dispersions (at $k_x = 0$) of four coupled cavity modes with the linear splitting $\alpha = 0$ (G) and $\alpha \neq 0$ (H).

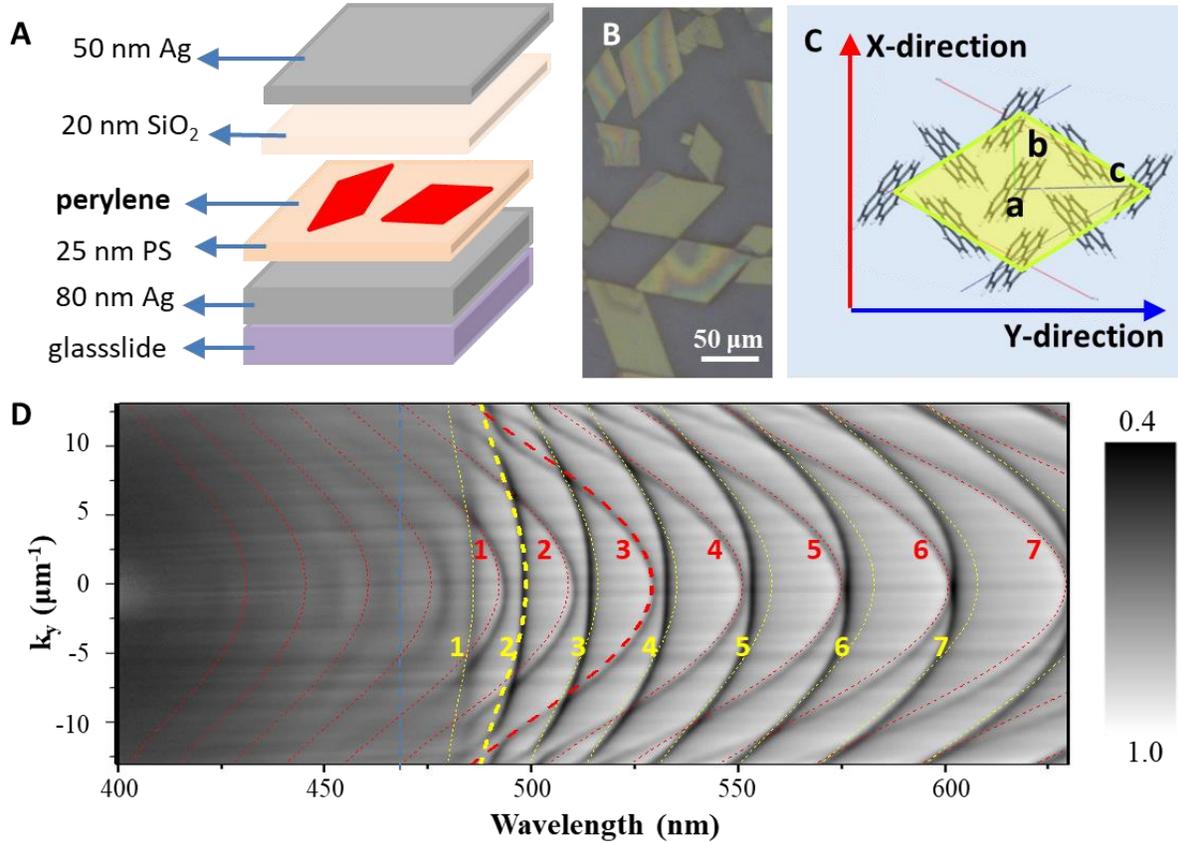

**Fig. 2. Organic single-crystalline microcavity and its angle-resolved reflectivity.** (A) Schematic diagram of the microcavity structure. (B) Microscopy image of the as-prepared β-phase perylene microcrystals. (C) Molecular orientation in the β-phase perylene crystal. Its unit cell comprises two molecules, forming a standard herringbone staking pattern. The two testing X- and Y-directions are denoted by the red and blue arrows, respectively. (D) Angle-resolved reflectivity of the microcavity with the organic-layer thickness of 3.2 μm. The red and yellow

dashed lines represent the simulation of the TE and TM cavity modes, respectively. The numbers are the corresponding indices.

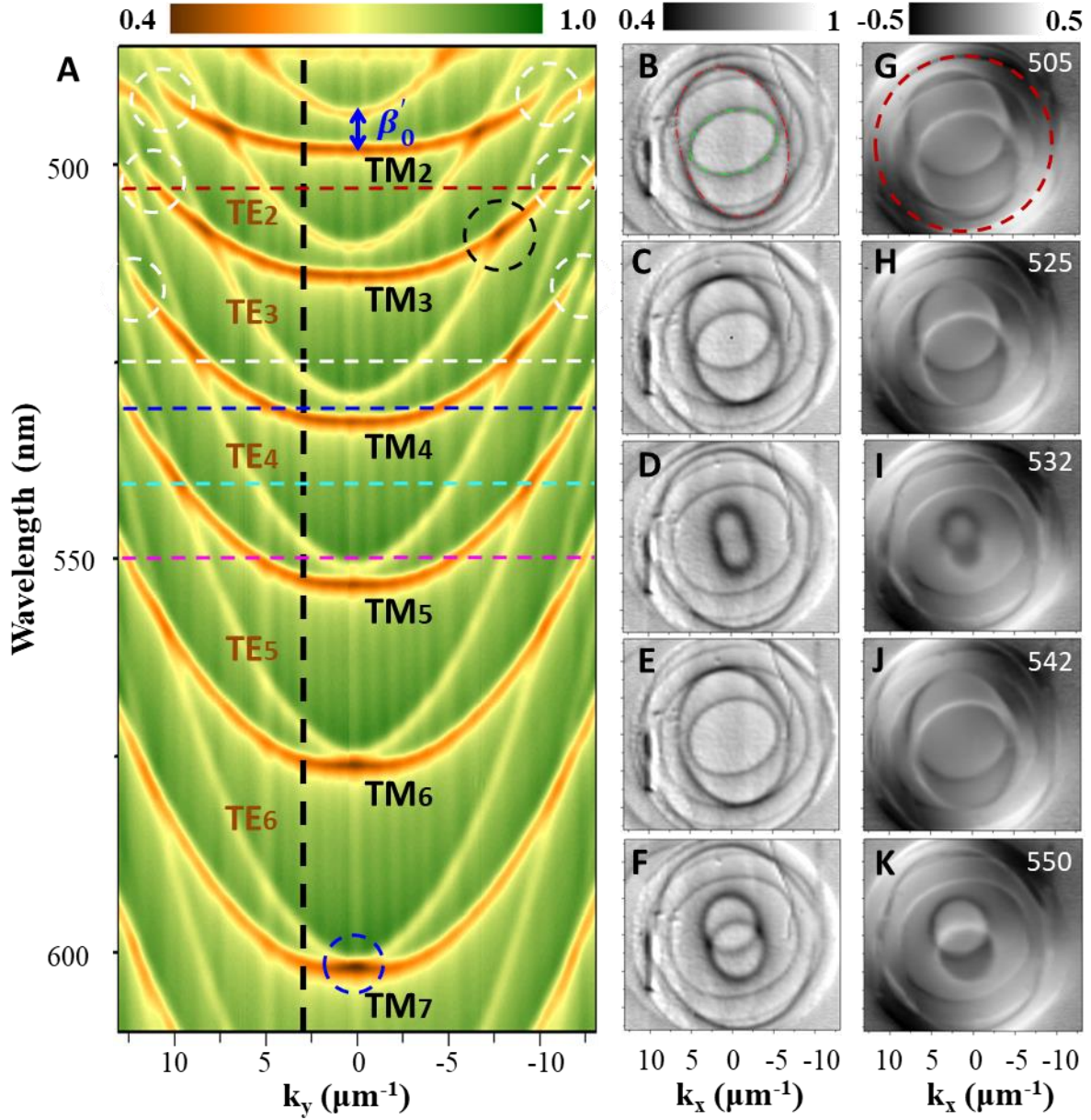

**Fig. 3. Dispersion relation and cavity modes.** (A) Magnified angle-resolved reflectivity of Fig. 2D. The black and white dash circles indicate the crossing and anti-crossing, respectively, of the TE and TM cavity modes. The subscript numbers represent the corresponding mode indices. The blue circle marks the crossing caused by the Rashba-Dresselhaus spin-orbit interaction. $\beta'_0$ represents the energy splitting of the $TM_n$ mode and the $TE_{n-1}$ mode. (B-F) Cross section maps of the 2D tomography at 505 nm (B), 525 nm (C), 532 nm (D), 542 nm (E), and 550 nm (F) in momentum space, corresponding to the horizontal dashed lines in (A), respectively. The red and green dashed circles in (B) represent the $TM_3$ mode and $TE_2$ mode, respectively. (G-K) Corresponding 2D wavevector maps of the Stokes vector $S_3$ component. The dashed circle in (G) represents the $TE_4$ mode.

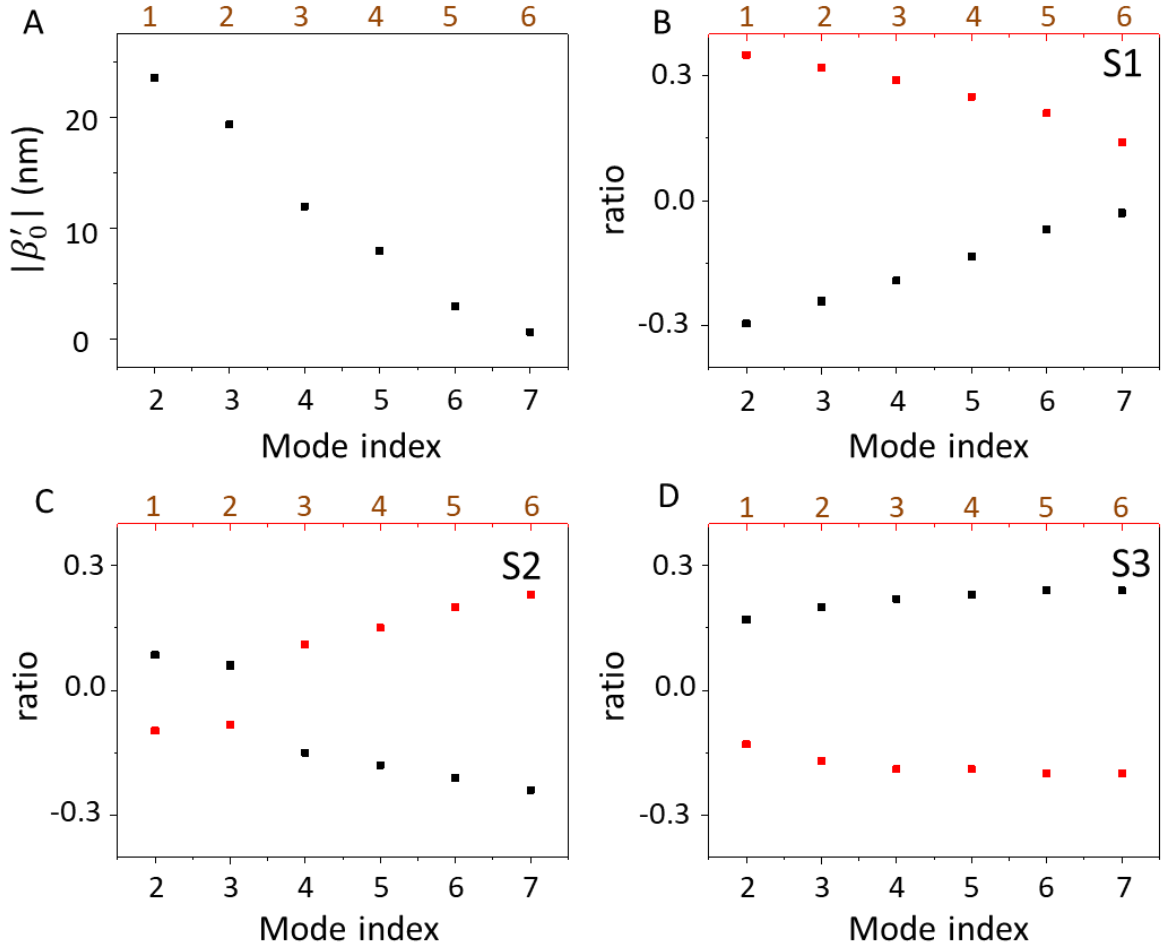

**Fig. 4. Wavelength-dependent energy splitting $\beta'_0$ and Stokes vectors.** (A) Dependence of the absolute value of the energy splitting $\beta'_0$ between the $TM_n$ mode and the $TE_{n-1}$ mode on mode indices. (B-D) Dependences of $S_1$, $S_2$, and $S_3$ components of the Stokes vectors on mode indices. In each panel, the upper indices denote the TE modes, while the lower indices denote to the TM modes. In (B-D), the red squares correspond to the upper label and the black squares correspond to the lower label.

# Supporting Information

**Realization of exciton-mediated optical spin-orbit interaction in organic microcrystalline resonators**

*Jiahuan Ren,[1,2] Qing Liao,[1,*] Xuekai Ma,[3,*] Stefan Schumacher,[3,4] Jiannian Yao,[2] Hongbing Fu[1,*]*

[1]Beijing Key Laboratory for Optical Materials and Photonic Devices, Department of Chemistry, Capital Normal University, Beijing 100048, People's Republic of China

[2] Tianjin Key Laboratory of Molecular Optoelectronic Science, School of Chemical Engineering and Technology, Tianjin University and Collaborative Innovation Center of Chemical Science and Engineering (Tianjin), Tianjin 300072, P. R. China

[3]Department of Physics and Center for Optoelectronics and Photonics Paderborn (CeOPP), Universität Paderborn, Warburger Strasse 100, 33098 Paderborn, Germany

[4]Wyant College of Optical Sciences, University of Arizona, Tucson, Arizona 85721, United States

## MATERIALS AND METHODS

**Preparation of the β-phase perylene single microcrystals**

Perylene (99+%) and TBAB (Tetrabutylammonium bromide) were purchased from Acros and Innochem, respectively. They were directly used without further purification.
The β-phase perylene nanosheets were prepared by the space-confined self-assembly method (Wang, Q.Yang, F. Zhang, Y. Chen, M. Zhang, X. Lei, S. Li, R. Hu, W. *J. Am. Chem. Soc.* **2018**, 140, 5339.). Briefly, 50 μL of 5 mg/ml perylene/chlorobenzene was slowly added into the 1mg/ml TBAB/water surface. After the chlorobenzene was evaporated, the rhombic β-phase perylene microcrystals were obtained.

**Preparation of the organic microcavities**

Firstly, an 80-nm silver film (reflectivity R ≥ 99% at 500 nm) was evaporated on a glass substrate, followed by spin-coating of a 25-nm polystyrene (PS) film. The as-prepared β-phase perylene microcrystals were transferred onto the PS film by bringing them in contact at the surface of the TBAB/water solution. A final evaporation of a 20-nm $SiO_2$ and a 50-nm silver film (R ≥ 90% at 500 nm) was made to form the microcavity.

**Assignment of the β-phase perylene single microcrystals**

To further confirm the crystal structure of the as-prepared rhombus perylene, the X-ray diffraction (XRD) measurements were carried out and shown in Figure S1. Two sharp peaks corresponding to the (100) and (200) crystal-face family for the β-phases perylene in the XRD pattern (Z. Z. Li, L. S. Liao, X. D. Wang, *Small* **2018**, 14, 1702952.) clearly indicates that the rhombus sheets are the β-phase crystal structure.

The transmission spectrum of the as-prepared rhombus perylene was measured and presented in Figure S2. The curves in black (red) colors denote the measurements with light polarized along b-axis (c-axis). The unique, polarization-independent peak around 473 nm in the experimental data indicates that the obtained perylene microcrystals are β-phase, which agrees with the previous reports (T. Rangel, A. Rinn, S. Sharifzadeh, F. da Jornada, A. Pick ,S. Louie, G. Witte, L Kroik, J. Neaton, S. Chatterjee, *PNAS* **2018**, 115, 284.).

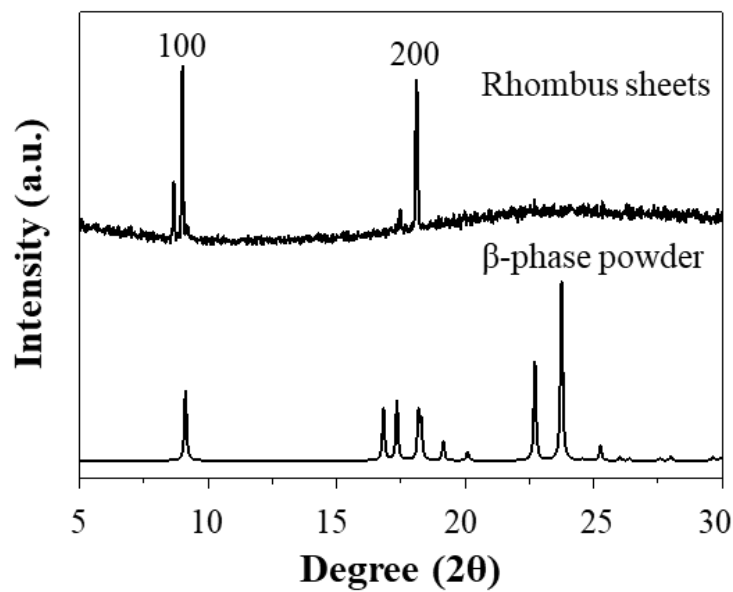

**Figure S1.** X-ray diffraction (XRD) of the β-phase perylene crystals (top) and powder (bottom). Two sharp peaks corresponding to the (100) and (200) crystal-face family for the β-phases perylene in the XRD pattern clearly indicates that the rhombus sheets are the β-phase crystal structure.

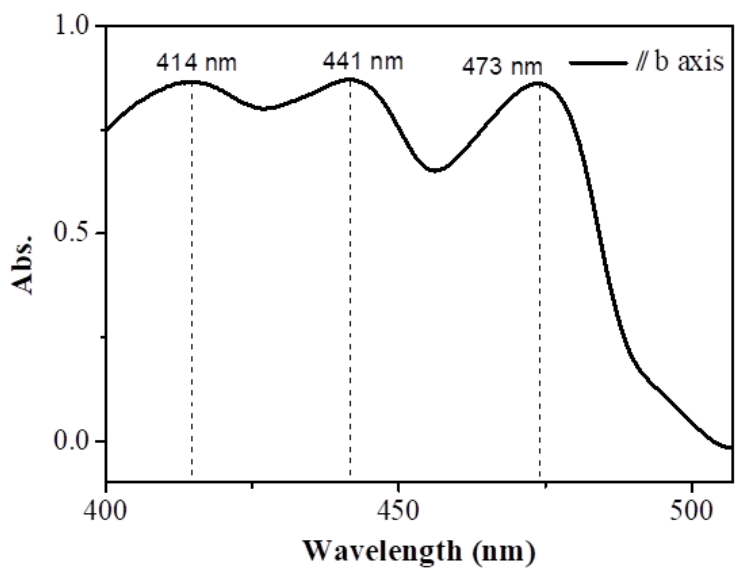

**Figure S2.** The transmission spectrum of the β-phase perylene crystals. The curves in black colors denote the measurements with light polarized along b-axis. The unique, polarization-independent peak around 473 nm in the experimental data indicates that the obtained perylene microcrystals are β-phase, which agrees with the previous reports.

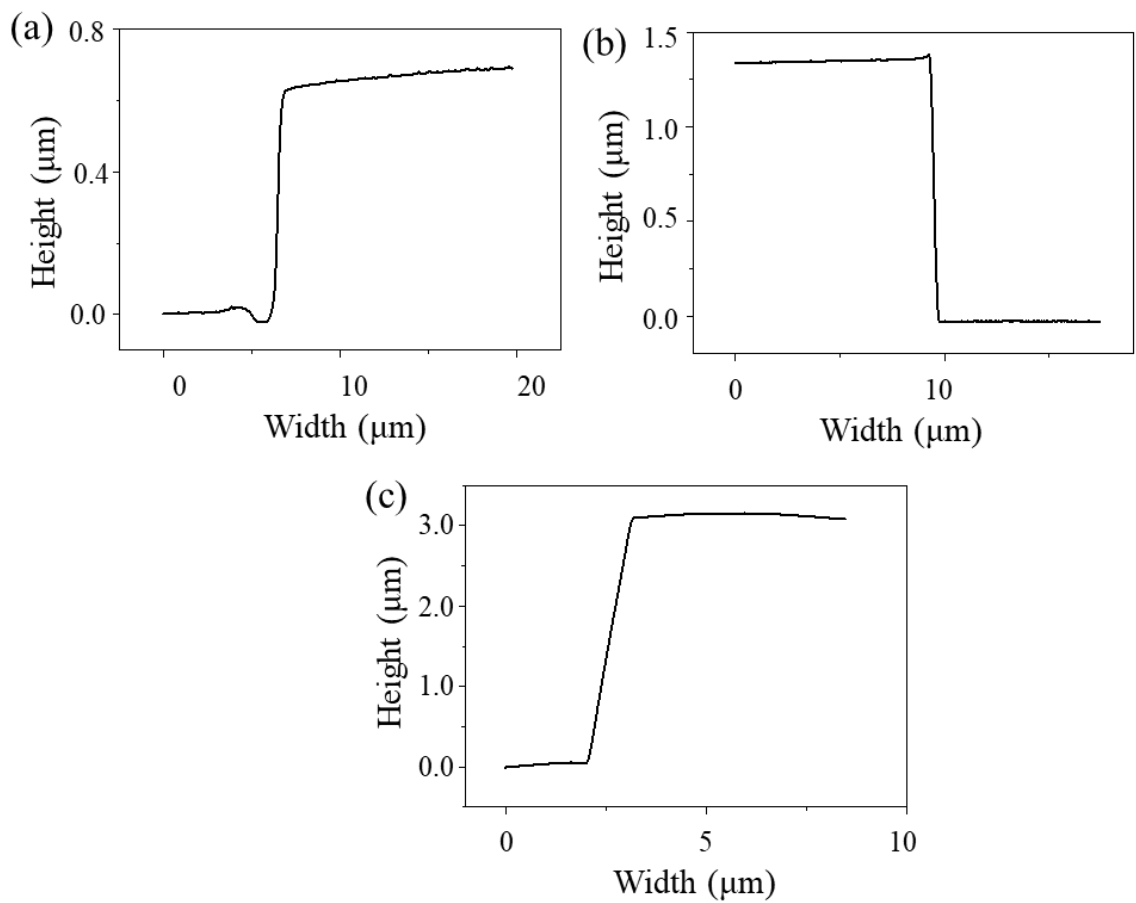

**Figure S3.** Atomic force microscope (AFM) of the β-phase perylene crystals. The thicknesses of these microcrystals were determined to be about 0.2 - 4 μm according to the AFM measurements. (a-c) The typical thicknesses of the β-phase perylene crystals.

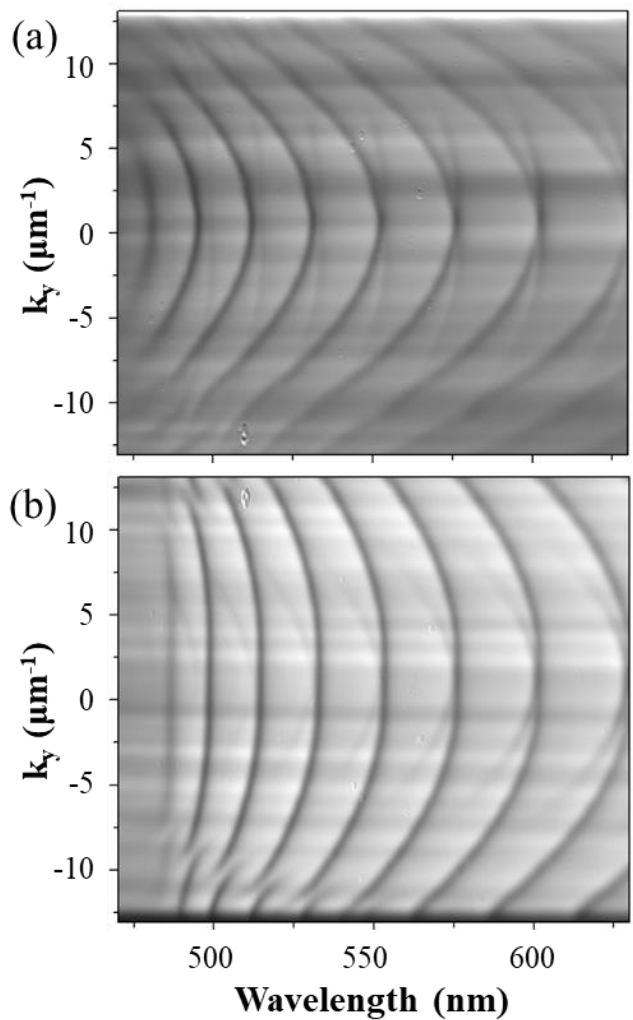

**Figure S4.** (a) Polarization-dependent angle-resolved reflectivity detected along X-direction, showing only the modes with larger curvatures. (b) Polarization-dependent angle-resolved reflectivity detected along Y-direction, showing only the modes with smaller curvatures.

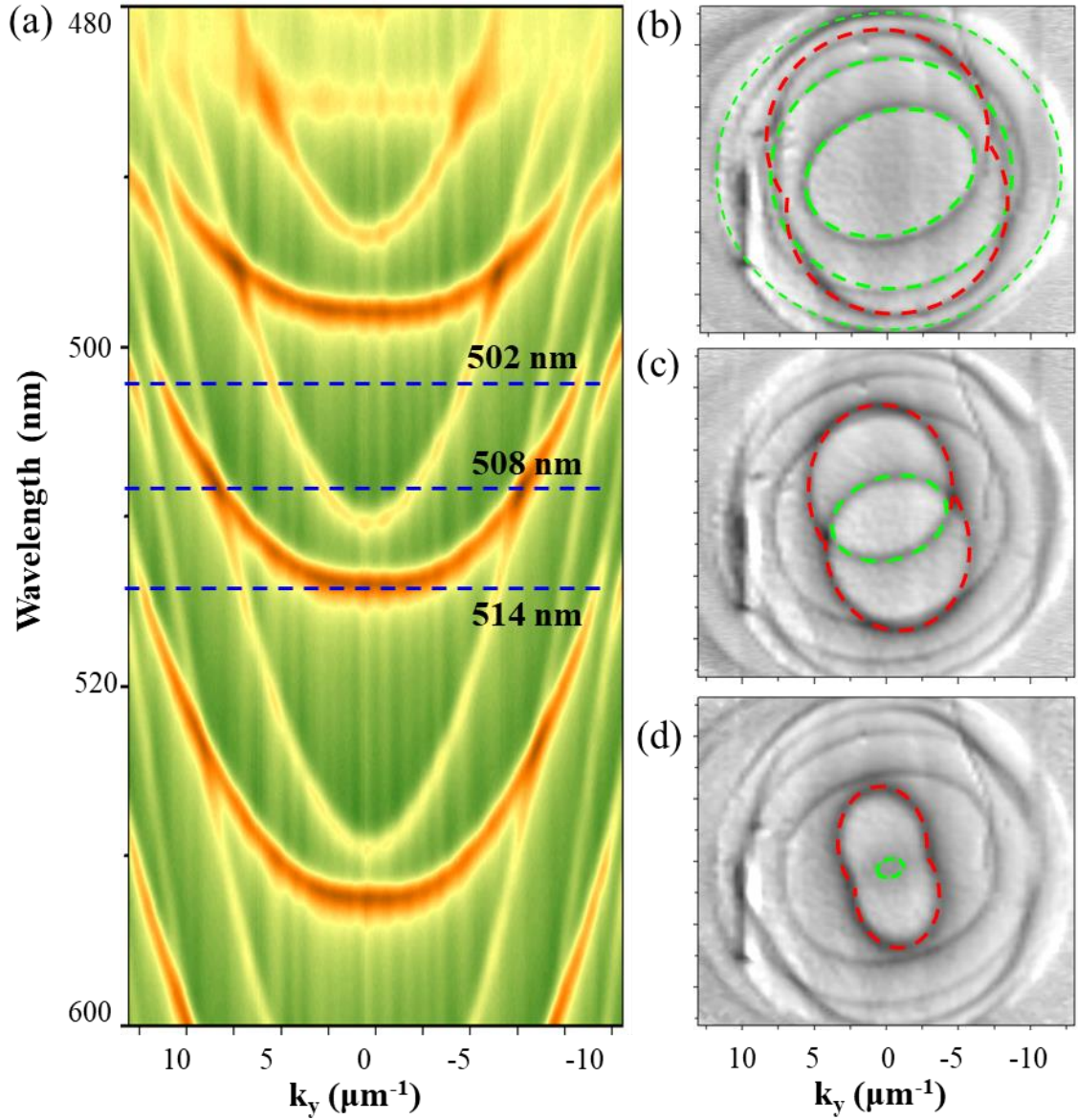

**Figure S5.** (a) The angle-resolved reflectivity at the wavelength ranging from 480 nm to 600 nm. The cross sections of 2D tomography at 502 nm (b), 508 nm (c) and 514 nm (d) in in-plane momentum space, respectively. Two sets of orthogonal ovals are observed, representing TE (green) and TM cavity modes (red), respectively.

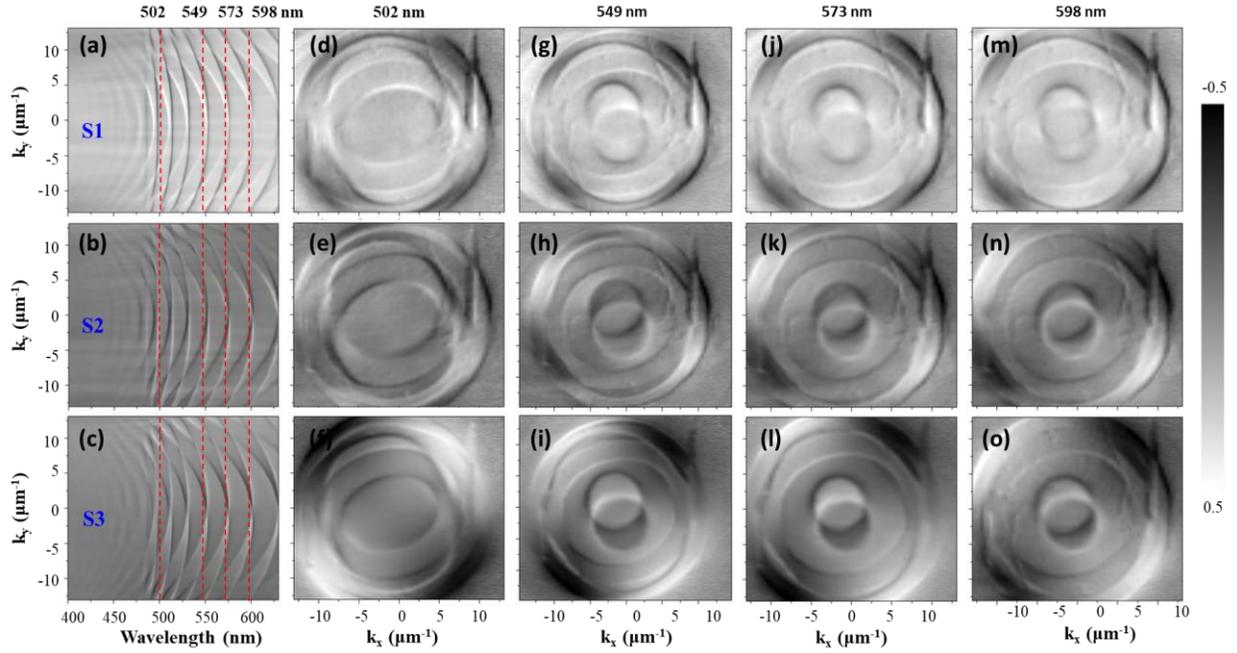

**Figure S6.** (a-c) The Stokes parameters $S_1$, $S_2$ and $S_3$ of the cavity modes in the β-perylene-filled microcavity. The corresponding 2D wavevector maps of the cross sections at 502 nm (d-f), 549 nm (g-i), 573 nm (j-l) and 598 nm (m-o), respectively.

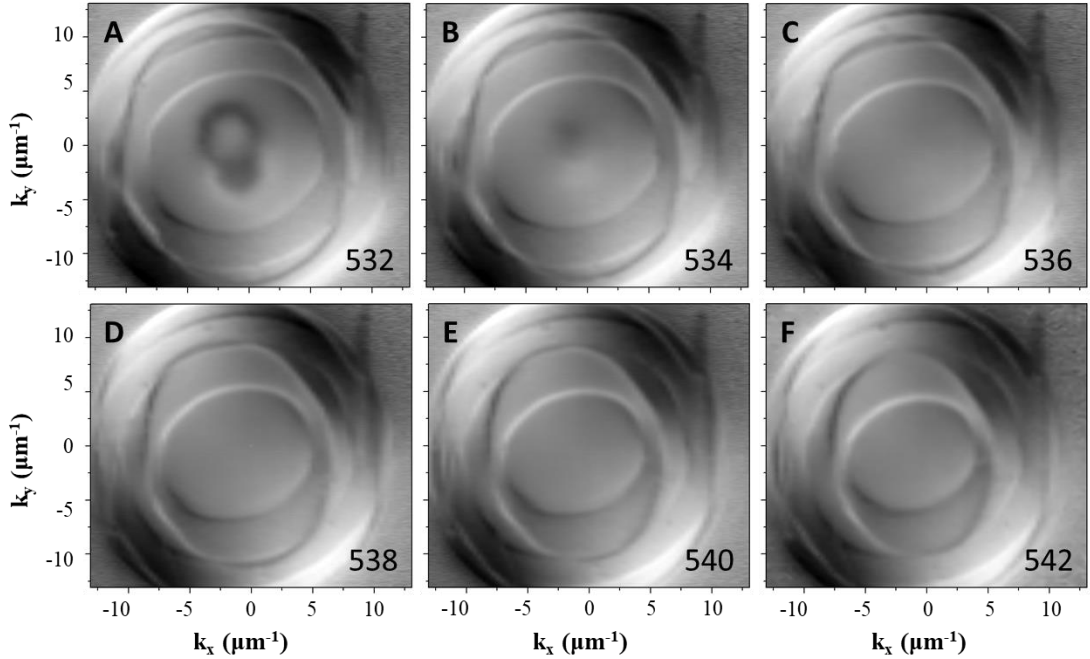

**Figure S7.** Cross sections ($S_3$ component) of the 2D tomography at 532 nm, 534 nm, 536 nm, 538 nm, 540 nm, and 542 nm in momentum space.